# Bidirectional acoustic negative refraction based on a pair of metasurfaces with both local and global *PT*-symmetries

Jun Lan,[1] Xiaowei Zhang,[1] Liwei Wang,[1] Yun Lai,[2,*] and Xiaozhou Liu[1,*]

[1]*Key Laboratory of Modern Acoustics, Institute of Acoustics and School of Physics, Collaborative Innovation Center of Advanced Microstructures, Nanjing University, Nanjing 210093, P. R. China*

[2]*Key Laboratory of Modern Acoustics, National Laboratory of Solid State Microstructures, School of Physics, and Collaborative Innovation Center of Advanced Microstructures, Nanjing University, Nanjing 210093, P. R. China*

Negative refraction plays an important role in acoustic wave manipulation and imaging. However, conventional systems based on acoustic metamaterials suffer from the limits induced by loss-related and resolution issues. In this work, a parity-time (*PT*)-symmetric system is introduced to realize loss-free bidirectional acoustic negative refraction. The system is composed of a pair of locally *PT*-symmetric multi-layer metasurfaces sandwiching a region of free space, which also forms a global *PT*-symmetry. The property of bidirectional negative refraction, which is rare for regular *PT*-symmetric structures, is related to the coexistence of amplification and absorption in the locally *PT*-symmetric metasurfaces at their *PT* broken phases. Such metasurfaces can freely switch their states between coherent perfect absorber (CPA) and amplifier depending on the direction of incidence. Our results provide a new physical mechanism for realizing bidirectional functions in *PT*-symmetric systems.

*Corresponding author: xzliu@nju.edu.cn (Xiaozhou Liu); laiyun@nju.edu.cn (Yun Lai)

## I. INTRODUCTION

Veselago used the constitutive parameters of dielectric permeability $\mu$ and magnetic permittivity $\varepsilon$ to define the concepts of single negative and double negative in electromagnetic metamaterial [1]. The unusual phenomenon of negative refraction can be induced by the simultaneously negative values of $\mu$ and $\varepsilon$ (i.e., double negativity (DNG)) [2,3], which promises a wide range of potential applications such as superlens [4,5] and illusion optics [6]. Since light and sound are both waves with similar wave characteristics, tremendous interests have been extracted to the acoustics analogue of DNG materials. Constitutive parameters for acoustic media are mass density $\rho$ and bulk modulus *B*, respectively [7,8]. If these two parameters are both negative, the phase velocity of acoustic wave is also negative and negative refraction can be realized. In previous studies, negative refraction has been realized by the artificial structures for classical acoustic waves, such as metamaterials and phononic crystals [9-11]. However, the property of negative refraction also inevitably leads to significantly increased sensitivity to losses, which imposes inherent challenges in practical applications.

In recent years, the investigation of electromagnetic *PT* symmetry has provided a new approach to realize negative refraction without the need of negative index materials, which may overcome the limitation of the DNG metamaterial designs [12-16]. The system of negative refraction is composed of a pair of single-layer metasurfaces exhibiting loss or gain, which enables unidirectional loss-free negative refraction. In this case, the power flow always directly transfers from the gain metasurface to the lossy metasurface, thus the negative refraction is always unidirectional, i.e., the negative refraction is possible only when the wave is incident on the lossy metasurface [17,18]. This unidirectional property comes from the exceptional point (EP) of the *PT*-symmetric system. In a flurry of researches, many extraordinary phenomena associated with the singular EP have been demonstrated, such as light-light switching [19,20], unidirectional invisibility [21-24], teleportation [25], and impurity-immunity [26]. At the EP, the eigenvalues and eigenvectors of the non-Hermitian system coalesce simultaneously [27]. Besides the EP, there exists another type of singular points denoted as the CPA-laser (CPAL) point in the electromagnetic *PT* symmetry, which has recently gained attention owing to its singular characteristics for the *PT*-symmetric system [28-32]. At the CPAL point, the eigenvalues go to either zero or infinity, corresponding to two mutually exclusive states, i.e., the coherent perfect absorption mode and the lasing mode.

Here, by introducing the concept of CPAL point into acoustic *PT* symmetry, similar CPA-saser (CPAS) point can be observed in the field of acoustics. Here, saser is the acoustic equivalence of optical laser [33]. Therefore, the acoustic *PT*-symmetric system can simultaneously behave as a perfect absorber that absorbs the incident acoustic waves, and an amplifier that amplifies the incident acoustic waves. Based on the characteristics of the CPAS point, a general method to realize bidirectional acoustic negative refraction in *PT*-symmetric systems is proposed in this work. The bidirectional negative refraction is realized by using two identical locally *PT*-symmetric multi-layer metasurfaces, separated by a region of free space. The loss and gain layers in the metasurfaces are carefully engineered to achieve the switching between the amplification and perfect absorption states at the CPAS point. Therefore, the negative refraction is independent of the incident direction of the acoustic waves, which is contrary to previous results in systems with a *PT* symmetry. Moreover, we find that this *PT*-symmetric systems can be designed to achieve negative bending effects by any desired angle, as well as planar focusing effects with good resolution.

## II. DESIGN

The bidirectional acoustic negative refraction system is sketched in Fig. 1, which is a globally *PT*-symmetric system constructed by two identical locally *PT*-symmetric multi-layer metasurfaces, which are separated by a region of free space with distance $d$. Each metasurface is composed of a four-layer structure with loss layers ($A$) and gain layers ($B$) arranged alternatively and periodically. The lengths of the loss and gain layers are both one-quarter wavelength $l$ = 25 mm and the total length of the metasurface is $L = \lambda_0$ ($\lambda_0$ is the wavelength in the air). Here, the operating frequency should be equal to the Bragg frequency $f_b$ = 3430 Hz and the CPAS point of the $PT$-symmetric metasurface is approached at this frequency. As shown in Fig. 1, when a plane wave is incident from the left side of $PT$-symmetric system with angle $\theta$ (black arrow in the I region), the multi-reflection between two metasurfaces (II region) forms both forward-traveling and backward-traveling acoustic waves. With suitable parameters, the forward-traveling wave can be ignored compared with the backward-traveling wave, and thus the forward-traveling wave is not shown in Fig. 1, as shall be explained later. In this sense, the main energy flux flows from the right side to the left side between two metasurfaces. When the backward-traveling wave transfers to the left metasurface, it will be absorbed by the left metasurface which is under the CPA mode, and there is no reflection in the I region. Meanwhile, when the forward-traveling wave transfers to the right metasurface, it will be transmitted after amplified by the right metasurface. The pressure amplitude of the transmitted wave in the III region is equal to that in the I region, which means the perfect transmission is achieved. Similarly, when a plane wave is incident from the right side of $PT$-symmetric system with angle $\theta$ (red arrow in the III region), the right metasurface absorbs waves from both sides and the left metasurface amplifies the incident wave from II region. Therefore, the bidirectional acoustic negative refraction functionality is obtained as a result of a pair of CPA and amplifier in such a $PT$-symmetric system. At the CPAS point, The $PT$-symmetric metasurface can uniquely satisfy both functionalities of the amplification and coherent perfect absorption without resorting to altering the frequency or the structure of the metasurface.

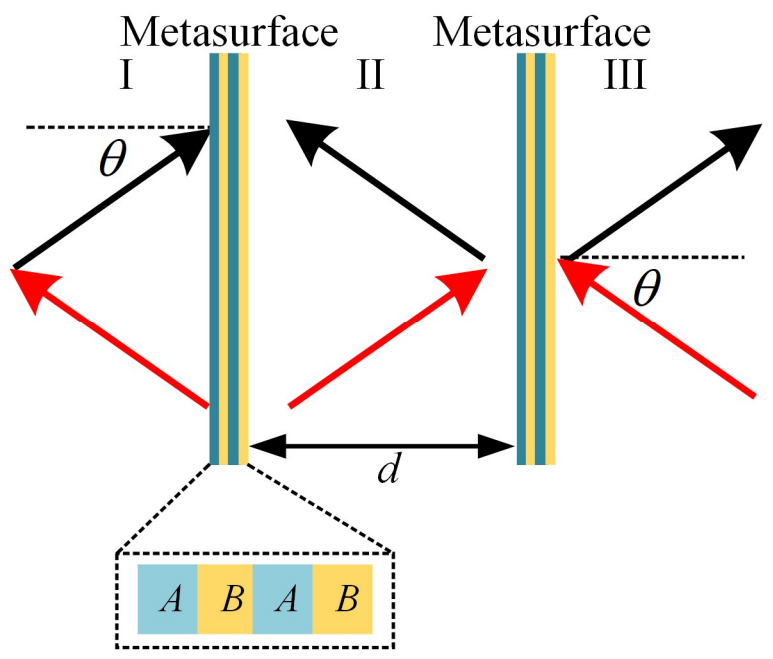

FIG. 1. The proposed bidirectional acoustic negative refraction system is composed of a pair of locally *PT*-symmetric multi-layer metasurfaces separated by a distance *d*. The black dotted box marks the arrangement of the loss (*A*) and gain (*B*) layers of the locally *PT*-symmetric metasurfaces.

## III. SCATTERING PROPERTIES

*PT* symmetry offers a new strategy to utilize loss to control gain, and the intriguing possibility to significantly expand the methods of acoustic wave manipulation [34]. Here, our *PT*-symmetric system is depicted in Fig. 1, which exhibits unique scattering property of bidirectional acoustic negative refraction. Since the metasurfaces exhibit local *PT* symmetry, the real part of the refractive index is symmetric, while the imaginary part of the refractive index is anti-symmetric. In our case, the refractive indices of the *A* and *B* layers are $n_A = n_0 + 0.5\delta i$ and $n_B = n_0 - 0.5\delta i$, respectively, where $n_0$ denotes the background refractive index ($n_0 = 1$) and $\delta$ denotes the loss-gain factor. The value of $\delta$ will be calculated later. The mass densities of the *A* and *B* layers are set as $\rho_A = \rho_B = 1.21\ \mathrm{kg/m^3}$ . For the bidirectional acoustic negative refraction system shown in Fig. 1, the scattering matrix *S* subject to such system with two ports can be expressed as

$$\begin{pmatrix} p_{bl} \\ p_{fr} \end{pmatrix} = S\begin{pmatrix} p_{br} \\ p_{fl} \end{pmatrix} = \begin{pmatrix} t_1 & r_L \\ r_R & t_1 \end{pmatrix}\begin{pmatrix} p_{br} \\ p_{fl} \end{pmatrix}, \tag{1}$$

where $p_{f(b)l}$ and $p_{f(b)r}$ are the components of the forward (backward)-traveling acoustic wave in the I and III regions, respectively. $r_{L(R)}$ is the reflection coefficient for left (right) incident acoustic wave. *t* is the transmission coefficient, which is identical for both left and right incident waves due to reciprocity. We start with a design for the case of normal incidence on the *PT*-symmetric system. The operating frequency is chosen as $f$ = 3400 Hz, which is slightly larger than expected Bragg frequency $f_b$. Based on the transfer matrix method in acoustics (see Appendix A), the magnitudes of transmission *t*, left reflection $r_L$, and right reflection $r_R$ coefficients as functions of the values of *d* and $\delta$ are calculated, as shown in Figs. 2(a)-2(c). It is noted that at the specific value of $\delta = 0.7$, the perfect transmission (i.e., $|t| = 1$) and non-reflection (i.e., $|r_L| = |r_R| = 0$) are obtained for the incidence from the left and right sides, which is independent of the distance *d* between two metasurfaces. In addition, the horizontal blue lines in Figs. 2(b) and 2(c) are induced by the Fabry-Perot resonance of the region of free space between two metasurfaces. At the resonant condition for standing-wave excitation ($d = (2n-1)\lambda_0/4$, where *n* is an integer), the transmission enhancement and zero reflection can be generated at the two ports of the system [35]. As a result, the proposed *PT*-symmetric system exhibits the bidirectional perfect transmission by appropriately selecting the value of $\delta$ or *d*. For the condition of $\delta = 0.7$ and arbitrary *d*, the corresponding scattering matrix of the globally *PT*-symmetric system describing the relation between the incoming and outgoing waves is given by

$$S = \begin{pmatrix} e^{i(k_0 d\cos\theta + \pi)} & 0 \\ 0 & e^{i(k_0 d\cos\theta + \pi)} \end{pmatrix}, \tag{2}$$

where $k_0$ is the wave number in free space and $\theta$ represents the incident angle. Zero reflection is obtained for the incidence from the left or right side of the system, and the transmitted wave undergoes a *phase advance* $(k_0 d\cos\theta + \pi)$ that is exactly opposite to the one without the pair of *PT*-symmetric metasurfaces (the time-harmonic dependence is $e^{i\omega t}$). Therefore, when $\delta = 0.7$, the designed *PT*-symmetric system exhibits loss-free zero reflection, and realizes bidirectional acoustic negative refraction just like a DNG medium of thickness *d* with an additional phase shift $\pi$. The additional phase shift $\pi$ is caused by the *PT*-symmetric metasurfaces. In this case, the scattering matrix describes the fascinating acoustic property of bidirectional acoustic negative refraction, which is rare for regular *PT*-symmetric systems.

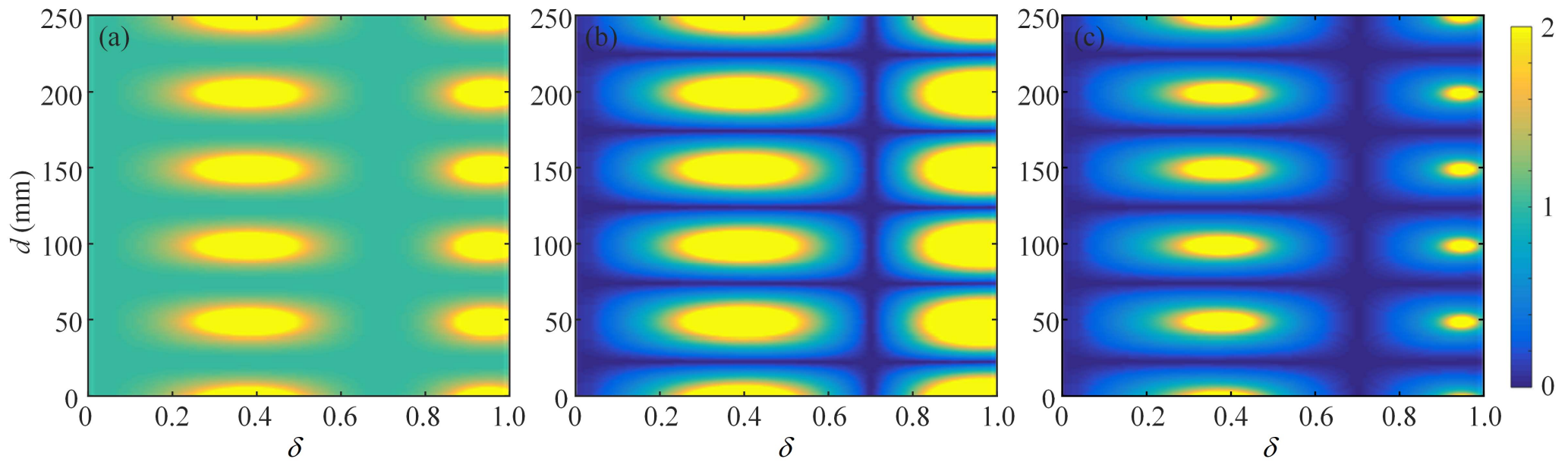

FIG. 2. Magnitudes of (a) transmission $t$, (b) left reflection $r_L$, and (c) right reflection $r_R$ coefficients of the globally $PT$-symmetric system as functions of the values of $d$ and $\delta$ for the normally incident wave at the operating frequency 3400 Hz.

The characteristics of the bidirectional reflectionless and phase advance have been guaranteed by Eq. (2), which theoretically confirm that the bidirectional acoustic negative refraction can be successfully achieved by a pair of locally $PT$-symmetric metasurfaces. The sound field distributions in the I and III regions shown in Fig. 1 are induced by an acoustic wave propagating from the metasurface far away from the side of incidence, to the metasurface close to the side of incidence, as represented by the black or red arrow (II region). When an acoustic wave is incident from the left side of $PT$-symmetric system, by using a transfer-matrix formalism, the acoustic waves on both I and II regions are connected through a transfer matrix

$$\begin{pmatrix} p_{fm} \\ p_{bm} \end{pmatrix} = T_l \begin{pmatrix} p_{fl} \\ p_{bl} \end{pmatrix}, \tag{3}$$

where $T_l$ and $p_{f(b)m}$ are the transfer matrix and the component of the forward (backward)-traveling acoustic wave in the II region, the details of which are given in Eq. (A9) of Appendix A. Similarly, when an acoustic wave is incident from the right side of $PT$-symmetric system, the acoustic waves on both II and III regions are connected through a transfer matrix

$$\begin{pmatrix} p_{fm} \\ p_{bm} \end{pmatrix} = T_r' \begin{pmatrix} p_{fr} \\ p_{br} \end{pmatrix}, \tag{4}$$

where $T_r'$ is the corresponding transfer matrix, the detail of which is given in Eq. (A12) of Appendix A. For the proposed $PT$-symmetric system, we have obtained that the absolute values of reflection wave $|p_{bl}|$ for left incidence in I region and reflection wave $|p_{fr}|$ for right incidence in III region are almost zero, as shown in Figs. 2(b) and 2(c). By substituting $|p_{bl}|=0$ and $|p_{fr}|=0$ into Eqs. (3) and (4), respectively, the absolute values of the forward-traveling components $|p_{fm}|$ for the acoustic waves incident from the left side and right side cases are $0.0013|p_{fl}|$ and $1.8|p_{br}|$, and the absolute values of the backward-traveling components $|p_{bm}|$ for the acoustic waves incident from left side and right side cases are $0.556|p_{fl}|$ and $0.0013|p_{br}|$. Therefore, the propagating direction of the total acoustic wave in II region is opposite to the direction of the incident acoustic wave in these two cases. Moreover, the relative phase difference between the left incidence in I region (or right incidence in III region) and the total acoustic wave in II region is around $-\pi/2\,(\pi/2)$.

This bidirectional property is related to the states of two locally *PT*-symmetric metasurfaces which can uniquely satisfy both the amplification and coherent perfect absorption states at the singular CPAS point. In the following, we demonstrate the states of two metasurfaces in the globally *PT*-symmetric system through analyzing the scattering property of the single metasurface. The black dotted box in Fig. 1 has shown the distribution of the loss and gain layers of the *PT*-symmetric metasurface. When $\delta=0.7$, the corresponding acoustic velocities in the loss and gain layers are $c_l=305.25+107.35i$ and $c_g=305.25-107.35i$, respectively. The transfer matrix method is used to derive the acoustic scattering matrix describing the relation between the input and output waves, i.e., $\begin{pmatrix} p_{O1} \\ p_{O2} \end{pmatrix} = S_1 \begin{pmatrix} p_{I2} \\ p_{I1} \end{pmatrix} = \begin{pmatrix} t_1 & r_{L1} \\ r_{R1} & t_1 \end{pmatrix} \begin{pmatrix} p_{I2} \\ p_{I1} \end{pmatrix}$, where $p_{I(O)1}$ and $p_{I(O)2}$ are the components of the input (output) waves at the left and right ports of the metasurface, respectively. $r_{L1(R1)}$ and $t_1$ are the left-(right-) reflection and transmission coefficients of the metasurface, respectively. The two eigenvalues of scattering matrix $S_1$ are expressed as $\lambda_{1,2}=t_1\pm\sqrt{r_{L1}r_{R1}}$. Figure 3(a) presents the absolute values of two eigenvalues $|\lambda_{1,2}|$ as a function of frequency for the locally *PT*-symmetric metasurface. It is seen that, at the operating frequency $f$ = 3400 Hz, the absolute values of two eigenvalues $|\lambda_{1,2}|$ go to either zero or infinity and the local metasurface is in the *PT* broken phase, which means that the metasurface approaches CPAS point. At such a point, the *PT*-symmetric metasurface simultaneously behaves as coherent amplifier and CPA. Moreover, Figs. 3(b) and 3(c) plot the amplitude and phase of the transmitted waves, the left- and right-reflected waves for the *PT*-symmetric metasurface with input amplitude 1 Pa, respectively. The results in Fig. 3(b) show that the magnitudes of the transmission, left reflection and right reflection coefficients are all large values ($|t_1| \approx 771$, $|r_{L1}| \approx 428.6$ and $|r_{R1}| \approx 1387$), indicating huge scattering is obtained at the CPAS point, which means that this metasurface could behave as an amplifier. Figure 3(c) shows that, at the CPAS point, there is a $\pi$ phase difference between two reflections, and the phase differences between the transmission and left- and right-reflections are $\mp\pi/2$. Therefore, in this *PT*-symmetric metasurface, coherent perfect absorption and amplification could be obtained. According to above discussions, the required conditions of the proposed *PT*-symmetric metasurface for coherent perfect absorption and amplification are given by

$$\begin{cases} t_1 + a r_{L1} e^{i\Delta\phi} = p_{O1}/p_{I2} \\ r_{R1} + a t_1 e^{i\Delta\phi} = p_{O2}/p_{I2} \end{cases}, \tag{5}$$

where $\Delta\phi$ is the relative phase difference between the left and right incident waves, $a = |p_{I1}|/|p_{I2}|$ is the absolute ratio of the left to right incident waves. Equation (5) indicates that, when $a \approx 1.8$ ($|t_1|/|r_{L1}| \approx |r_{R1}|/|t_1| = 1.8$), the coherent perfect absorption is achieved for $\Delta\phi = -\pi/2$. Conversely, the coherent amplification is achieved for $\Delta\phi = \pi/2$. In addition, we have obtained the pressure field of the forward- and backward-traveling waves in the I, II and III regions of the globally *PT*-symmetric system. When an acoustic wave is incident from the left side of *PT*-symmetric system, for the left metasurface, the ratio of the absolute values of the left to right incident waves (backward-traveling component $p_{bm}$) is $|p_{fl}|/|p_{bm}| \approx 1.8 (= a)$, and the relative phase difference between the left and right incident waves is $-\pi/2$. Thus, the left metasurface acts as a CPA. Meanwhile, the right metasurface behaves as an amplifier. When an incident wave with absolute pressure $|p_{fm}| = 0.0013|p_{fl}|$ is incident on the right metasurface, the absolute pressure of the transmitted wave in the III region is equal to that of the incident wave in the I region, i.e., $|p_{fr}| = |p_{fm}||t_1| = |p_{fl}|$. Besides, when an acoustic wave is incident from the right side of *PT*-symmetric system, the right metasurface acts as a CPA and the left metasurface acts as an amplifier. Therefore, the pair of metasurfaces freely switches the states between CPA-amplifier pair and amplifier-CPA pair dependent on the direction of incidence.

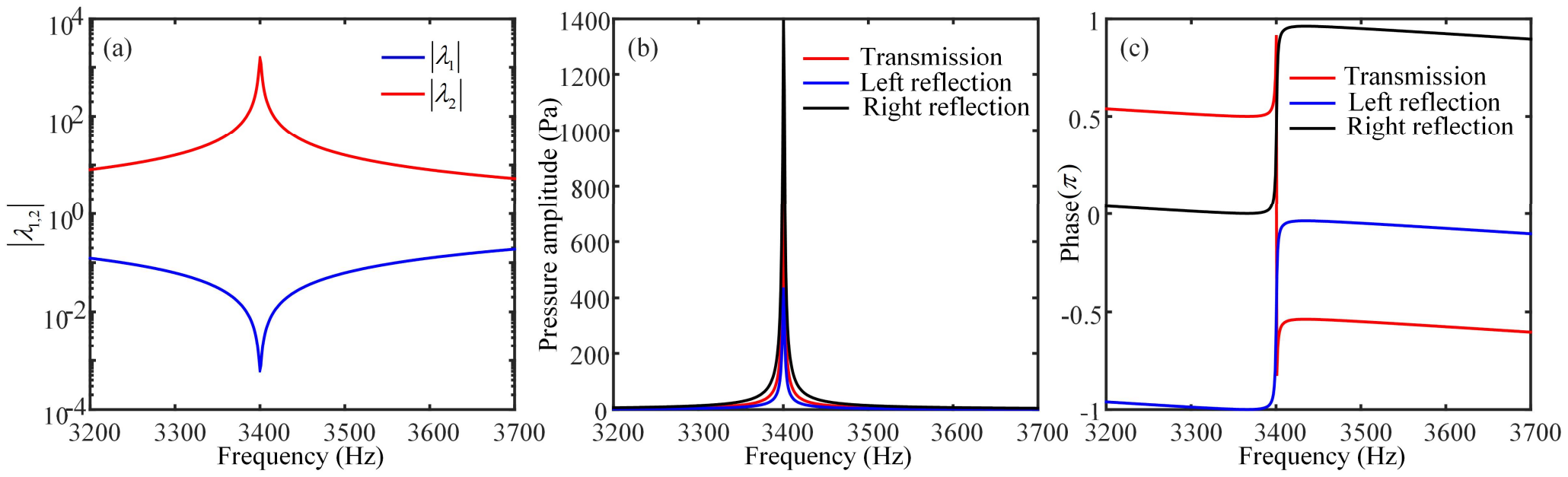

FIG. 3. (a) The absolute values of two eigenvalues $|\lambda_{1,2}|$ as a function of frequency for the locally *PT*-symmetric metasurface. The magnitudes and phases of the transmission, left-reflection and right-reflection coefficients for the *PT*-symmetric metasurface are shown in (b) and (c), respectively.

## IV. BIDIRECTIONAL ACOUSTIC NEGATIVE REFRACTION AND PLANAR FOCUSING

To gain physical insight into this intriguing phenomenon, we have performed full-wave simulation by using a finite element solver (COMSOL Multiphysics software) to verify the bidirectional acoustic negative refraction effect of the designed *PT*-symmetric system. The plane acoustic wave is incoming with frequency located at the CPAS point of the *PT*-symmetric metasurface (i.e., $f = 3400$ Hz). Figures 4(a) and 4(b) show the simulated pressure field distributions of the systems under plane waves incident normally from the left and right sides, respectively, where the distance between two metasurfaces is $d$ = 500 mm for both cases. The white arrows in figures represent the direction of the power flow. In the numerical simulations, the front and back boundaries are perfect absorbing boundaries. As predicted by the theoretical analysis, the system is non-reflecting for the acoustic wave incident from the left and right sides, and the transmission coefficients are all around $|t| \approx 1$. This indicates that the pair of metasurfaces has unique ability to switch the states between CPA-amplifier pair and amplifier-CPA pair flexibility according to the direction of incident wave. Since energy flux flows from the amplifier to the CPA, the acoustic wave between two metasurfaces are propagating in the direction opposite to that of the incident wave, which provides a *phase advance* to the transmitted wave. Moreover, in Figs. 4(c) and 4(d), we replace the distance $d$ = 125 mm and still obtain left- (right-) reflection coefficient $r_{L(R)} \to 0$ and transmission coefficient $t \approx 1$, which indicates that the perfect transmission is independent of the distance $d$ between two metasurfaces. Hence, the acoustic perfect transmission is clearly observed for the normally incident plane wave and the globally $PT$-symmetric system has potential ability in bidirectional acoustic negative refraction.

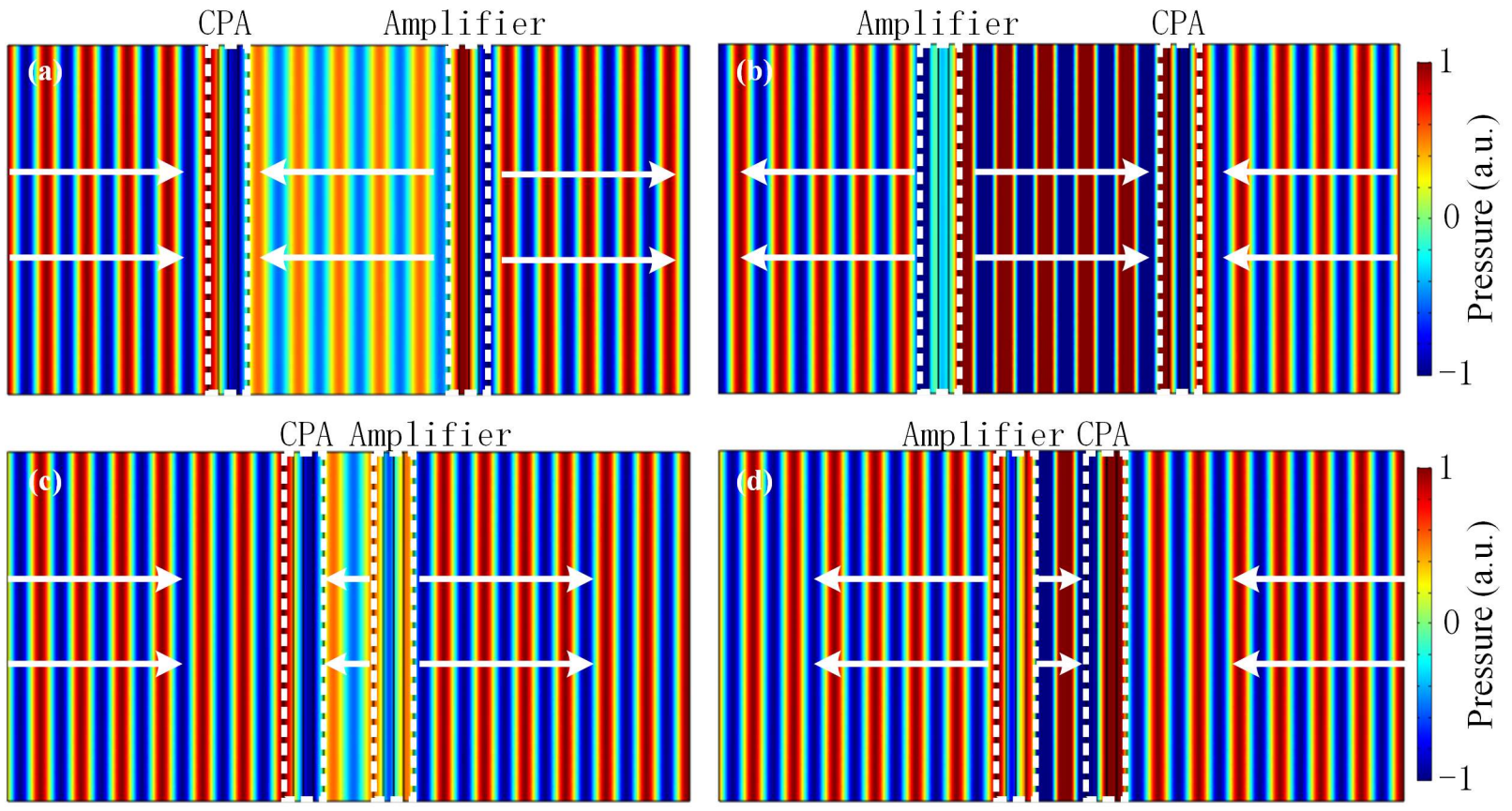

FIG. 4. Pressure field distributions of the bidirectional acoustic negative refraction system for the normally incident plane wave at the operating frequency of $f$ = 3400 Hz. The distances between two metasurfaces are set to (a), (b) $d$ = 500 mm and (c), (d) $d$ = 125 mm, respectively. In (a) and (c), the plane wave is incident from the left side. In (b) and (d), the plane wave is incident from the right side.

It is important to note that this $PT$-symmetric system can be designed to negatively bend acoustic incident waves with any desired angle, as shown in Eq. (2). For the obliquely incident wave with angle $\theta$, to obtain the bidirectional negative refraction at the operating frequency 3400 Hz, the acoustic densities and velocities of the loss and gain layers are set as $\rho_l(\theta)=\rho_l/\cos\theta$ , $\rho_g(\theta)=\rho_g/\cos\theta$ , $c_l(\theta)=c_l\cos\theta_l$ and $c_g(\theta)=c_g\cos\theta_g$ , where $\theta_l=\arcsin(c_l(\theta)/c_0\sin\theta)$ and $\theta_g=\arcsin(c_g(\theta)/c_0\sin\theta)$ are the refracted angles in loss and gain layers, respectively. For $\theta=25^{\circ}$ and $\theta=60^{\circ}$, the pressure field distributions are presented in Figs. 5(a), 5(b) and Figs. 5(c), 5(d), respectively. In Figs. 5(a) and 5(c), the plane acoustic wave is incident from the left side. However, in Figs. 5(b) and 5(d), the plane acoustic wave is incident from the right side. Here, the distance between two metasurfaces is set as $d$ = 500 mm for both cases. Clearly, even for obliquely incident condition, the perfect transmission still exists in the system. The bidirectional acoustic negative refraction is clearly observed between two metasurfaces, where power transfers from the amplifier to the CPA. At CPAS point, the pair of *PT*-symmetric metasurfaces still has unique ability to switch states between CPA-amplifier pair and amplifier-CPA pair simply by tuning the incident direction. Hence, this *PT*-symmetric system supports bidirectional acoustic negative refraction for obliquely incident wave.

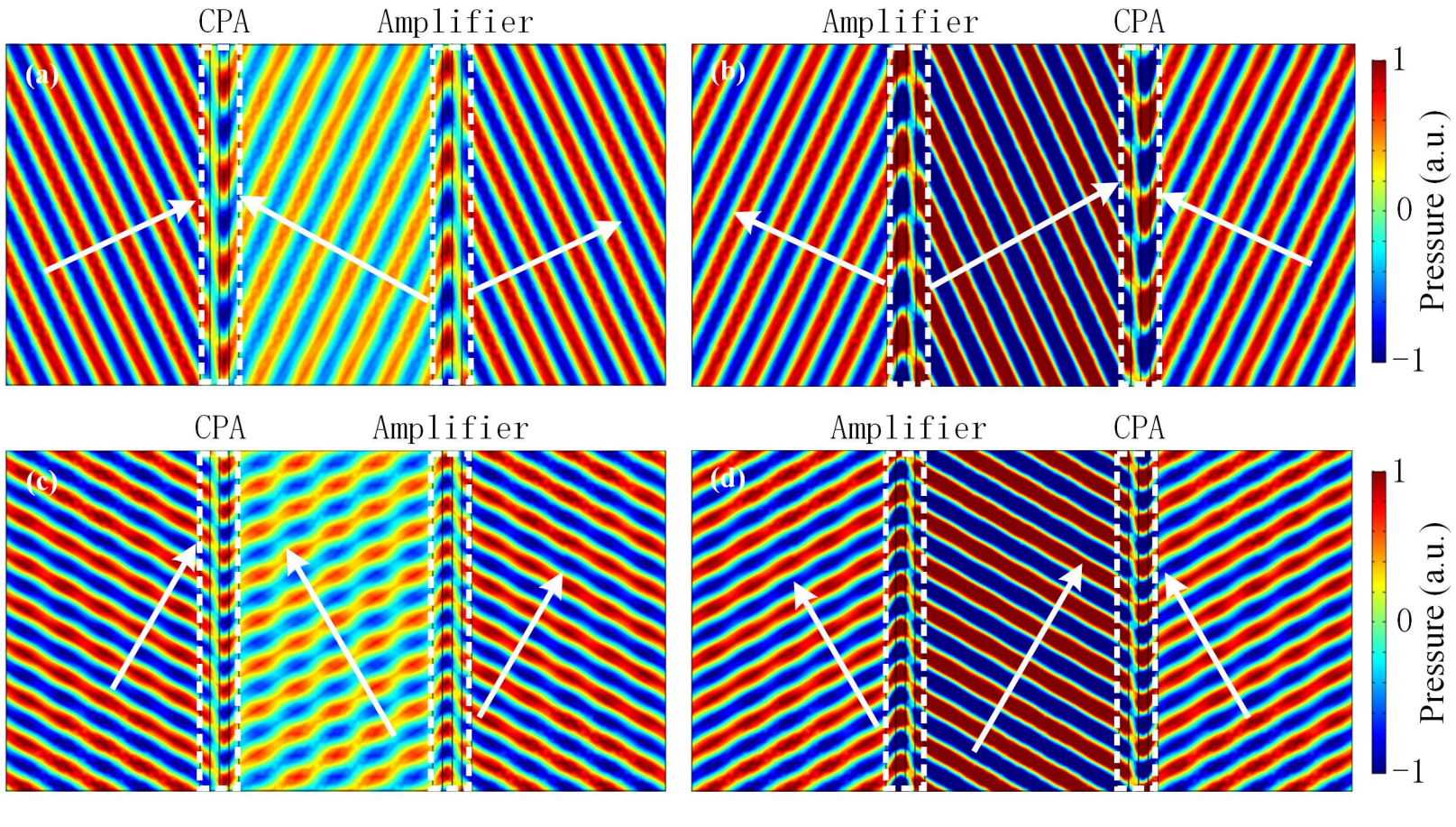

FIG. 5. Pressure field distributions of the bidirectional acoustic negative refraction system for (a), (b) $\theta=25^{\circ}$ and (c), (d) $\theta=60^{\circ}$ at the operating frequency 3400 Hz. The white arrows represent the direction of the power flow. In (a) and (c), the plane wave is incident from the left side. In (b) and (d), the plane wave is incident from the right side.

Since the *PT*-symmetric system has good resolution in both transverse and longitudinal directions, it has potential in ideally focusing the source with arbitrary angle [13]. If the incidence is a point source at a distance from the *PT*-symmetric system, it will again focus at two different distance points, which generates two images of the point source. Here, we assume the point source is placed at a distance $d_1 = 25l$ from the system on the left or right side, and the distance between two *PT*-symmetric metasurfaces is $2d_1$. The acoustic densities and velocities of the loss and gain layers along the longitudinal direction are variable, which are expressed as $\rho_l(\theta) = \rho_l / \cos\theta$, $\rho_g(\theta) = \rho_g / \cos\theta$, $c_l(\theta) = c_l \cos\theta_l$ and $c_g(\theta) = c_g \cos\theta_g$, where $\theta = \arctan(y/d_1)$ is dependent on the position in *y*-axis. Figures 6(a) and 6(b) show the simulated pressure field distributions of the proposed *PT*-symmetric system under point sources located on the left and right sides, respectively. As expected, two images of the point source are created at the center of two metasurfaces and the outside region of the system, respectively. Moreover, Fig. 6(c) shows the pressure intensity field on the focus plane at $d_1$ distance from the *PT*-symmetry system. The half-power beam widths of the focus spots for both conditions of the point sources located on the left and right sides are all around $0.5\lambda_0$, which confirms that the proposed *PT*-symmetric system ideally focuses the propagating spectrum and offers good resolution. Therefore, the bidirectional *PT*-symmetric planar focusing is realized by the proposed *PT*-symmetric system.

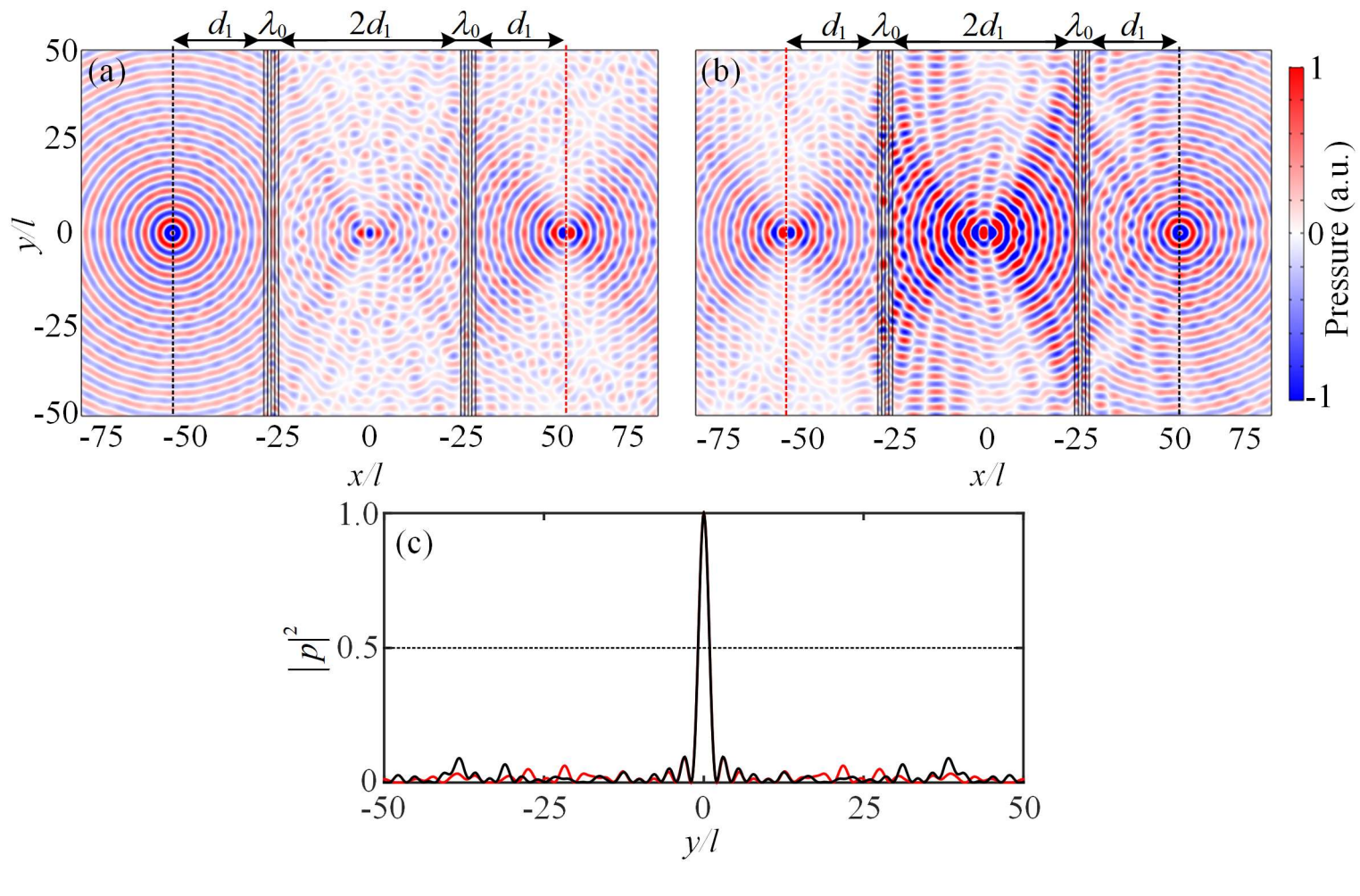

FIG. 6. Pressure field distributions of the bidirectional acoustic planar focusing. The incident point sources are located on the left (a) and right (b) sides of the *PT*-symmetric system, respectively. (c) Red and black lines represent the intensity field on the focus planes shown in (a) and (b) with red dotted lines, respectively.

## V. CONCLUSION

In conclusion, we have proposed a concept and theory to achieve bidirectional acoustic negative refraction without the need of DNG metamaterials. A new type of CPAS points at which a *PT*-symmetric multi-layer acoustic metasurface operates as a CPAS is revealed. A pair of locally *PT*-symmetric metasurfaces that sandwich a region of free space is used to design a globally *PT*-symmetric system. At the singular CPAS point, this pair of metasurfaces can flexibly switch the states between CPA-amplifier pair and amplifier-CPA pair without resorting to altering the frequency or the structure. This feature provides an approach to realizing bidirectional negative refraction with perfect transmission and planar focusing with good resolution, different from the previous unidirectional behaviors realized in simple *PT*-symmetric systems. Our results may stimulate the potential of bidirectional characteristics in *PT*-symmetric systems and further prove the validity of using *PT*-symmetry as a new method to overcome the limitations of passive system.

## ACKNOWLEDGEMENTS

This work was supported by the National Key R&D Program (No. 2017YFA0303702), State Key Program of National Natural Science Foundation of China (No. 11834008), National Natural Science Foundation of China (No. 11774167, No. 61671314, and No. 11974176), State Key Laboratory of Acoustics, Chinese Academy of Science (No. SKLA201809), Key Laboratory of Underwater Acoustic Environment, Chinese Academy of Sciences (No. SSHJ-KFKT-1701) and AQSIQ Technology R&D Program (No. 2017QK125).

## APPENDIX A: TRANSFER MATRIX METHOD

In this Appendix, the globally *PT*-symmetric system is analyzed by the transfer matrix method. As shown in Fig. A1, the *PT*-symmetric system is composed of eleven sections: four loss layers ($A$), four gain layers ($B$), and three free space regions (I, II, and III regions). When a plane acoustic wave is incident on the system, it will generate multiple transmissions and reflections that bound forth and back within each section. In Fig. A1, we show the stationary forward- and backward-traveling waves in each section, which are the sum of an infinite number of transient forward- and backward-traveling waves, respectively. When a plane acoustic wave is incident from the left side of the *PT*-symmetric system. The expressions of the pressure field ( $p_{\mathrm{I}}$ ) and the particle velocity field ( $v_{\mathrm{I}}$ ) that contain forward- and backward-traveling waves in the I region ( $x \leq 0$ ) are given by

$$p_{\mathrm{I}} = p_{fl} e^{jk_0(x\cos\theta_f - y\sin\theta_f)} + p_{bl} e^{-jk_0(x\cos\theta_b + y\sin\theta_b)}, \quad \text{(A1)}$$

$$v_{\mathrm{I}} = \frac{\cos\theta_f}{\rho_0 c_0} p_{fl} e^{jk_0(x\cos\theta_f - y\sin\theta_f)} - \frac{\cos\theta_b}{\rho_0 c_0} p_{bl} e^{-jk_0(x\cos\theta_b + y\sin\theta_b)}, \quad \text{(A2)}$$

where $\rho_0$, $c_0$ and $k_0$ are the density, the speed and the wave number in free space. $p_{fl}$ and $p_{bl}$ are the components of the forward- and backward-traveling acoustic waves in the I region, respectively. $\theta_f$ is the angle between forward traveling wave and *x*-axis (incident angle), and $\theta_b$ is the angle between backward traveling wave and *x*-axis (reflected angle). The pressure field and the associated particle velocity field that contain forward- and backward-traveling acoustic waves in the *m*-th layer of the left/right metasurface ((1, *m*)/(2, *m*), *m* = 1,2,3,4) are given by

$$p_{(1,m)} = p_f^{(1,m)} e^{-jk^{(1,m)}((x-ml)\cos\theta_f^{(1,m)} - y\sin\theta_f^{(1,m)})} + p_b^{(1,m)} e^{jk^{(1,m)}((x-ml)\cos\theta_b^{(1,m)} + y\sin\cos\theta_b^{(1,m)})}, \quad \text{(A3)}$$

$$v_{(1,m)} = \frac{\cos\theta_f^{(1,m)}}{\rho^{(1,m)} c^{(1,m)}} p_f^{(1,m)} e^{-jk^{(1,m)}((x-ml)\cos\theta_f^{(1,m)} - y\sin\theta_f^{(1,m)})} - \frac{\cos\theta_b^{(1,m)}}{\rho^{(1,m)} c^{(1,m)}} p_b^{(1,m)} e^{jk^{(1,m)}((x-ml)\cos\theta_b^{(1,m)} + y\sin\theta_b^{(1,m)})}, \quad \text{(A4)}$$

$$p_{(2,m)} = p_f^{(2,m)} e^{-jk^{(2,m)}((x-(4+m)l-d)\cos\theta_f^{(2,m)} - y\sin\theta_f^{(2,m)})} + p_b^{(2,m)} e^{jk^{(2,m)}((x-(4+m)l-d)\cos\theta_b^{(2,m)} + y\sin\theta_b^{(2,m)})}, \quad \text{(A5)}$$

$$v_{(2,m)} = \frac{\cos\theta_f^{(2,m)}}{\rho^{(2,m)} c^{(2,m)}} p_f^{(2,m)} e^{-jk^{(2,m)}((x-(4+m)l-d)\cos\theta_f^{(2,m)} - y\sin\theta_f^{(2,m)})} - \frac{\cos\theta_b^{(2,m)}}{\rho^{(2,m)} c^{(2,m)}} p_b^{(2,m)} e^{jk^{(2,m)}((x-(4+m)l-d)\cos\theta_b^{(2,m)} + y\sin\theta_b^{(2,m)})}, \quad \text{(A6)}$$

where $p_{(1,m)}$ and $v_{(1,m)}$ are the pressure and associated particle velocity fields in the $m$-th layer of the left metasurface, $p_{(2,m)}$ and $v_{(2,m)}$ are that in the $m$-th region of the right metasurface. $\rho^{(1,m)}(\rho^{(2,m)})$, $c^{(1,m)}(c^{(2,m)})$ and $k^{(1,m)}(k^{(2,m)})$ are the density, the speed and the wave number in the $m$-th layer of the left (right) metasurface. $\theta_f^{(1,m)}$ or $\theta_f^{(2,m)}$ is the angle between forward-traveling wave and $x$-axis in the $m$-th layer of the left or right metasurface. $\theta_b^{(1,m)}$ or $\theta_b^{(2,m)}$ is the angle between backward-traveling wave and $x$-axis in the $m$-th layer of the left or right metasurface. The pressure field ($p_{\text{II}}$) and particle velocity field ($v_{\text{II}}$) that contain forward- and backward-traveling waves in the II region ($4l<x<4l+d$) are given by

$$p_{\text{II}} = p_{fm}e^{-jk_0((x-4l)\cos\theta_f - y\sin\theta_f)} + p_{bm}e^{jk_0((x-4l)\cos\theta_b + y\sin\theta_b)}, \tag{A7}$$

$$v_{\text{II}} = \frac{\cos\theta_f}{\rho_0 c_0} p_{fm}e^{-jk_0((x-4l)\cos\theta_f - y\sin\theta_f)} - \frac{\cos\theta_b}{\rho_0 c_0} p_{bm}e^{jk_0((x-4l)\cos\theta_b + y\sin\theta_b)}. \tag{A8}$$

When a plane acoustic wave is incident from the left side of the $PT$-symmetric system, According to boundary conditions, the pressures and particle velocities in the interface of two different media are continuous, the forward- and backward-traveling acoustic waves at both sides of the left metasurface are connected through the transfer matrix, which can be expressed as

$$\begin{pmatrix} p_{fm} \\ p_{bm} \end{pmatrix} = T_l \begin{pmatrix} p_{fl} \\ p_{bl} \end{pmatrix}, \tag{A9}$$

where $T_l = T_{l4}T_{l3}T_{l2}T_{l1}T_{l0}$, in which

$$T_{l0} = \begin{bmatrix} \left(1+\frac{\rho^{(1,1)}c^{(1,1)}}{\rho_0 c_0}\frac{\cos\theta_f}{\cos\theta_f^{(1,1)}}\right)\Big/2 & \left(1-\frac{\rho^{(1,1)}c^{(1,1)}}{\rho_0 c_0}\frac{\cos\theta_f}{\cos\theta_f^{(1,1)}}\right)\Big/2 \\ \left(1-\frac{\rho^{(1,1)}c^{(1,1)}}{\rho_0 c_0}\frac{\cos\theta_f}{\cos\theta_f^{(1,1)}}\right)\Big/2 & \left(1+\frac{\rho^{(1,1)}c^{(1,1)}}{\rho_0 c_0}\frac{\cos\theta_f}{\cos\theta_f^{(1,1)}}\right)\Big/2 \end{bmatrix},$$

$$T_{lm}=M_{lm}N_{lm}=\begin{bmatrix}\left(1+\dfrac{\rho^{(1,m+1)}c^{(1,m+1)}}{\rho^{(1,m)}c^{(1,m)}}\dfrac{\cos\theta_f^{(1,m)}}{\cos\theta_f^{(1,m+1)}}\right)\Big/2 & \left(1-\dfrac{\rho^{(1,m+1)}c^{(1,m+1)}}{\rho^{(1,m)}c^{(1,m)}}\dfrac{\cos\theta_f^{(1,m)}}{\cos\theta_f^{(1,m+1)}}\right)\Big/2\\ \left(1-\dfrac{\rho^{(1,m+1)}c^{(1,m+1)}}{\rho^{(1,m)}c^{(1,m)}}\dfrac{\cos\theta_f^{(1,m)}}{\cos\theta_f^{(1,m+1)}}\right)\Big/2 & \left(1+\dfrac{\rho^{(1,m+1)}c^{(1,m+1)}}{\rho^{(1,m)}c^{(1,m)}}\dfrac{\cos\theta_f^{(1,m)}}{\cos\theta_f^{(1,m+1)}}\right)\Big/2\end{bmatrix}$$
$$\times\begin{bmatrix}e^{jk^{(1,m)}l\cos\theta_f^{(1,m)}} & 0\\ 0 & e^{-jk^{(1,m)}l\cos\theta_f^{(1,m)}}\end{bmatrix}\quad m=1,2,3$$

$$T_{l4}=M_{l4}N_{l4}=\begin{bmatrix}\left(1+\dfrac{\rho_0c_0}{\rho^{(1,4)}c^{(1,4)}}\dfrac{\cos\theta_f^{(1,4)}}{\cos\theta_f}\right)\Big/2 & \left(1-\dfrac{\rho_0c_0}{\rho^{(1,4)}c^{(1,4)}}\dfrac{\cos\theta_f^{(1,4)}}{\cos\theta_f}\right)\Big/2\\ \left(1-\dfrac{\rho_0c_0}{\rho^{(1,4)}c^{(1,4)}}\dfrac{\cos\theta^{(1,4)}}{\cos\theta_f}\right)\Big/2 & \left(1+\dfrac{\rho_0c_0}{\rho^{(1,4)}c^{(1,4)}}\dfrac{\cos\theta_f^{(1,4)}}{\cos\theta_f}\right)\Big/2\end{bmatrix}$$
$$\times\begin{bmatrix}e^{jk^{(1,4)}l\cos\theta_f^{(1,4)}} & 0\\ 0 & e^{-jk^{(1,4)}l\cos\theta_f^{(1,4)}}\end{bmatrix}.$$

It should be noted that, according to Snell's law, the angle of the forward-traveling wave is equal to the one of the backward-traveling wave in each section, i.e., $\theta_f^{(1,m)}=\theta_b^{(1,m)}$, $\theta_f^{(2,m)}=\theta_b^{(2,m)}$ and $\theta_f=\theta_b$. The forward- and backward-traveling acoustic waves at both sides of the right metasurface are connected through the transfer matrix

$$\begin{pmatrix}p_{fr}\\ p_{br}\end{pmatrix}=T_r\begin{pmatrix}p_{fm}\\ p_{bm}\end{pmatrix}, \tag{A10}$$

where $T_r=T_{r4}T_{r3}T_{r2}T_{r1}T_{r0}$, in which

$$T_{r0}=M_{r0}N_{r0}=\begin{bmatrix}\left(1+\dfrac{\rho^{(2,1)}c^{(2,1)}}{\rho_0c_0}\dfrac{\cos\theta_f}{\cos\theta_f^{(1,2)}}\right)\Big/2 & \left(1-\dfrac{\rho^{(2,1)}c^{(2,1)}}{\rho_0c_0}\dfrac{\cos\theta_f}{\cos\theta_f^{(2,1)}}\right)\Big/2\\ \left(1-\dfrac{\rho^{(2,1)}c^{(2,1)}}{\rho_0c_0}\dfrac{\cos\theta_f}{\cos\theta_f^{(2,1)}}\right)\Big/2 & \left(1+\dfrac{\rho^{(2,1)}c^{(2,1)}}{\rho_0c_0}\dfrac{\cos\theta_f}{\cos\theta_f^{(2,1)}}\right)\Big/2\end{bmatrix}$$
$$\times\begin{bmatrix}e^{jk_0d\cos\theta_f} & 0\\ 0 & e^{-jk_0d\cos\theta_f}\end{bmatrix},$$

$$T_{rm} = M_{rm}N_{rm}$$
$$= \begin{bmatrix} \left(1+\frac{\rho^{(2,m+1)}c^{(2,m+1)}}{\rho^{(2,m)}c^{(2,m)}}\frac{\cos\theta_f^{(2,m)}}{\cos\theta_f^{(2,m+1)}}\right)\Big/2 & \left(1-\frac{\rho^{(2,m+1)}c^{(2,m+1)}}{\rho^{(2,m)}c^{(2,m)}}\frac{\cos\theta_f^{(2,m)}}{\cos\theta_f^{(2,m+1)}}\right)\Big/2 \\ \left(1-\frac{\rho^{(2,m+1)}c^{(2,m+1)}}{\rho^{(2,m)}c^{(2,m)}}\frac{\cos\theta_f^{(2,m)}}{\cos\theta_f^{(2,m+1)}}\right)\Big/2 & \left(1+\frac{\rho^{(2,m+1)}c^{(2,m+1)}}{\rho^{(2,m)}c^{(2,m)}}\frac{\cos\theta_f^{(2,m)}}{\cos\theta_f^{(2,m+1)}}\right)\Big/2 \end{bmatrix}$$
$$\times \begin{bmatrix} e^{jk^{(2,m)}l\cos\theta_f^{(2,m)}} & 0 \\ 0 & e^{-jk^{(2,m)}l\cos\theta_f^{(2,m)}} \end{bmatrix}, \quad m = 1,2,3$$

$$T_{r4} = M_{r4}N_{r4}$$
$$= \begin{bmatrix} \left(1+\frac{\rho_0 c_0}{\rho^{(2,4)}c^{(2,4)}}\frac{\cos\theta_f^{(2,4)}}{\cos\theta_f}\right)\Big/2 & \left(1-\frac{\rho_0 c_0}{\rho^{(2,4)}c^{(2,4)}}\frac{\cos\theta_f^{(2,4)}}{\cos\theta_f}\right)\Big/2 \\ \left(1-\frac{\rho_0 c_0}{\rho^{(2,4)}c^{(2,4)}}\frac{\cos\theta_f^{(2,4)}}{\cos\theta_f}\right)\Big/2 & \left(1+\frac{\rho_0 c_0}{\rho^{(2,4)}c^{(2,4)}}\frac{\cos\theta_f^{(2,4)}}{\cos\theta_f}\right)\Big/2 \end{bmatrix} \begin{bmatrix} e^{jk^{(2,4)}l\cos\theta_f^{(2,4)}} & 0 \\ 0 & e^{-jk^{(2,4)}l\cos\theta_f^{(2,4)}} \end{bmatrix}.$$

Therefore, when a plane acoustic wave is incident from the left side of the globally *PT*-symmetric system, the forward- and backward-traveling acoustic waves at both sides of the *PT*-symmetric system are connected through the transfer matrix

$$\begin{pmatrix} p_{fr} \\ p_{br} \end{pmatrix} = T_r T_l \begin{pmatrix} p_{fl} \\ p_{bl} \end{pmatrix} = T_{r4}T_{r3}T_{r2}T_{r1}T_{r0}T_{l4}T_{l3}T_{l2}T_{l1}T_{l0} \begin{pmatrix} p_{fl} \\ p_{bl} \end{pmatrix}. \qquad \text{(A11)}$$

In addition, when an acoustic wave is incident from the right side of the globally *PT*-symmetric system, the forward- and backward-traveling acoustic waves at both sides of the *PT*-symmetric system are connected through the transfer matrix

$$\begin{pmatrix} p_{fl} \\ p_{bl} \end{pmatrix} = T_l' T_r' \begin{pmatrix} p_{fr} \\ p_{br} \end{pmatrix} = T_{l1}'T_{l2}'T_{l3}'T_{l4}'T_{l0}'T_{r1}'T_{r2}'T_{r3}'T_{r4}'T_{r0}' \begin{pmatrix} p_{fr} \\ p_{br} \end{pmatrix}, \qquad \text{(A12)}$$

where $\begin{pmatrix} p_{fm} \\ p_{bm} \end{pmatrix} = T_r' \begin{pmatrix} p_{fr} \\ p_{br} \end{pmatrix}$, $\begin{pmatrix} p_{fl} \\ p_{bl} \end{pmatrix} = T_l' \begin{pmatrix} p_{fm} \\ p_{bm} \end{pmatrix}$, $T_r' = T_{r1}'T_{r2}'T_{r3}'T_{r4}'T_{r0}'$, and $T_l' = T_{l1}'T_{l2}'T_{l3}'T_{l4}'T_{l0}'$, in which

$$T_{r0}' = \begin{bmatrix} \left(1+\frac{\rho^{(2,4)}c^{(2,4)}}{\rho_0 c_0}\frac{\cos\theta_f}{\cos\theta_f^{(2,4)}}\right)\Big/2 & \left(1-\frac{\rho^{(2,4)}c^{(2,4)}}{\rho_0 c_0}\frac{\cos\theta_f}{\cos\theta_f^{(2,4)}}\right)\Big/2 \\ \left(1-\frac{\rho^{(2,4)}c^{(2,4)}}{\rho_0 c_0}\frac{\cos\theta_f}{\cos\theta_f^{(2,4)}}\right)\Big/2 & \left(1+\frac{\rho^{(2,4)}c^{(2,4)}}{\rho_0 c_0}\frac{\cos\theta_f}{\cos\theta_f^{(2,4)}}\right)\Big/2 \end{bmatrix},$$

$$T'_{r(m+1)} = M'_{r(m+1)} N'_{r(m+1)}$$

$$= \begin{bmatrix} \left(1 + \frac{\rho^{(2,m)} c^{(2,m)}}{\rho^{(2,m+1)} c^{(2,m+1)}} \frac{\cos\theta_f^{(2,m+1)}}{\cos\theta_f^{(2,m)}}\right) \Big/ 2 & \left(1 - \frac{\rho^{(2,m)} c^{(2,m)}}{\rho^{(2,m+1)} c^{(2,m+1)}} \frac{\cos\theta_f^{(2,m+1)}}{\cos\theta_f^{(2,m)}}\right) \Big/ 2 \\ \left(1 - \frac{\rho^{(2,m)} c^{(2,m)}}{\rho^{(2,m+1)} c^{(2,m+1)}} \frac{\cos\theta_f^{(2,m+1)}}{\cos\theta_f^{(2,m)}}\right) \Big/ 2 & \left(1 + \frac{\rho^{(2,m)} c^{(2,m)}}{\rho^{(2,m+1)} c^{(2,m+1)}} \frac{\cos\theta_f^{(2,m+1)}}{\cos\theta_f^{(2,m)}}\right) \Big/ 2 \end{bmatrix}$$

$$\times \begin{bmatrix} e^{jk^{(2,m+1)} l \cos\theta_f^{(2,m+1)}} & 0 \\ 0 & e^{-jk^{(2,m+1)} l \cos\theta_f^{(2,m+1)}} \end{bmatrix}, \quad m = 3,2,1$$

$$T'_{r1} = M'_{r1} N'_{r1}$$

$$= \begin{bmatrix} \left(1 + \frac{\rho_0 c_0}{\rho^{(2,1)} c^{(2,1)}} \frac{\cos\theta_f^{(2,1)}}{\cos\theta_f}\right) \Big/ 2 & \left(1 - \frac{\rho_0 c_0}{\rho^{(2,1)} c^{(2,1)}} \frac{\cos\theta_f^{(2,1)}}{\cos\theta_f}\right) \Big/ 2 \\ \left(1 - \frac{\rho_0 c_0}{\rho^{(2,1)} c^{(2,1)}} \frac{\cos\theta_f^{(2,1)}}{\cos\theta_f}\right) \Big/ 2 & \left(1 + \frac{\rho_0 c_0}{\rho^{(2,1)} c^{(2,1)}} \frac{\cos\theta_f^{(2,1)}}{\cos\theta_f}\right) \Big/ 2 \end{bmatrix} \begin{bmatrix} e^{jk^{(2,1)} l \cos\theta_f^{(2,1)}} & 0 \\ 0 & e^{-jk^{(2,1)} l \cos\theta_f^{(2,1)}} \end{bmatrix},$$

$$T'_{l0} = M'_{l0} N'_{l0} = \begin{bmatrix} \left(1 + \frac{\rho^{(1,4)} c^{(1,4)}}{\rho_0 c_0} \frac{\cos\theta_f}{\cos\theta_f^{(1,4)}}\right) \Big/ 2 & \left(1 - \frac{\rho^{(1,4)} c^{(1,4)}}{\rho_0 c_0} \frac{\cos\theta_f}{\cos\theta_f^{(1,4)}}\right) \Big/ 2 \\ \left(1 - \frac{\rho^{(1,4)} c^{(1,4)}}{\rho_0 c_0} \frac{\cos\theta_f}{\cos\theta_f^{(1,4)}}\right) \Big/ 2 & \left(1 + \frac{\rho^{(1,4)} c^{(1,4)}}{\rho_0 c_0} \frac{\cos\theta_f}{\cos\theta_f^{(1,4)}}\right) \Big/ 2 \end{bmatrix}$$

$$\times \begin{bmatrix} e^{jk_0 d \cos\theta_f} & 0 \\ 0 & e^{-jk_0 d \cos\theta_f} \end{bmatrix},$$

$$T'_{l(m+1)} = M'_{l(m+1)} N'_{l(m+1)}$$

$$= \begin{bmatrix} \left(1 + \frac{\rho^{(1,m)} c^{(1,m)}}{\rho^{(1,m+1)} c^{(1,m+1)}} \frac{\cos\theta_f^{(1,m+1)}}{\cos\theta_f^{(1,m)}}\right) \Big/ 2 & \left(1 - \frac{\rho^{(1,m)} c^{(1,m)}}{\rho^{(1,m+1)} c^{(1,m+1)}} \frac{\cos\theta_f^{(1,m+1)}}{\cos\theta_f^{(1,m)}}\right) \Big/ 2 \\ \left(1 - \frac{\rho^{(1,m)} c^{(1,m)}}{\rho^{(1,m+1)} c^{(1,m+1)}} \frac{\cos\theta_f^{(1,m+1)}}{\cos\theta_f^{(1,m)}}\right) \Big/ 2 & \left(1 + \frac{\rho^{(1,m)} c^{(1,m)}}{\rho^{(1,m+1)} c^{(1,m+1)}} \frac{\cos\theta_f^{(1,m+1)}}{\cos\theta_f^{(1,m)}}\right) \Big/ 2 \end{bmatrix}$$

$$\times \begin{bmatrix} e^{jk^{(1,m+1)} l \cos\theta_f^{(1,m+1)}} & 0 \\ 0 & e^{-jk^{(1,m+1)} l \cos\theta_f^{(1,m+1)}} \end{bmatrix}, \quad m = 3,2,1$$

$$T'_{l1} = M'_{l1} N'_{l1}$$

$$= \begin{bmatrix} \left(1 + \frac{\rho_0 c_0}{\rho^{(1,1)} c^{(1,1)}} \frac{\cos\theta_f^{(1,1)}}{\cos\theta_f}\right) \Big/ 2 & \left(1 - \frac{\rho_0 c_0}{\rho^{(1,1)} c^{(1,1)}} \frac{\cos\theta_f^{(1,1)}}{\cos\theta_f}\right) \Big/ 2 \\ \left(1 - \frac{\rho_0 c_0}{\rho^{(1,1)} c^{(1,1)}} \frac{\cos\theta_f^{(1,1)}}{\cos\theta_f}\right) \Big/ 2 & \left(1 + \frac{\rho_0 c_0}{\rho^{(1,1)} c^{(1,1)}} \frac{\cos\theta_f^{(1,1)}}{\cos\theta_f}\right) \Big/ 2 \end{bmatrix} \begin{bmatrix} e^{jk^{(1,1)} l \cos\theta_f^{(1,1)}} & 0 \\ 0 & e^{-jk^{(1,1)} l \cos\theta_f^{(1,1)}} \end{bmatrix}.$$

The reflection coefficients for the acoustic waves incident from the left side ($r_L$) and the right side ($r_R$) of the *PT*-symmetric system can be obtained from Eqs. (A11) and (A12), respectively. The transmission coefficients $t$ calculated from Eqs. (A11) and (A12) are the same due to reciprocity.

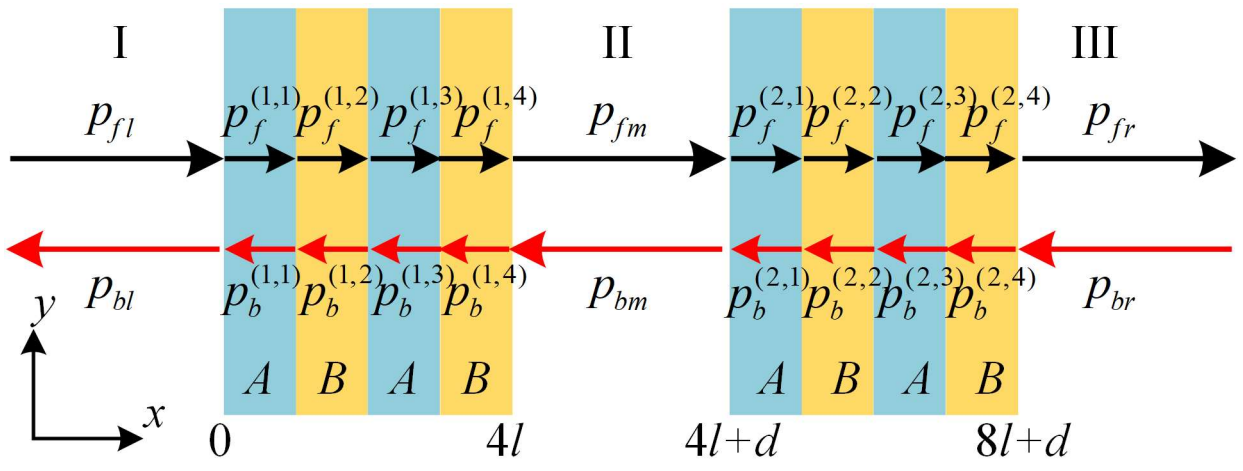

FIG. A1 Transfer matrix model of the globally *PT*-symmetric system.